\documentclass[twocolumn, graphics, floatfix, a4paper, aps, prb, superscriptaddress, longbibliography, showpacs]{revtex4-2}
\usepackage{graphicx}
\usepackage{bm}
\usepackage[usenames,dvipsnames]{xcolor}
\usepackage[colorlinks=true,urlcolor=blue,citecolor=blue,linkcolor=blue]{hyperref}
\usepackage{amssymb,amsmath}
\usepackage[normalem]{ulem}
\usepackage{booktabs}
\usepackage{longtable}
\usepackage{braket}
\usepackage[T1]{fontenc}
\usepackage[utf8]{inputenc}
\usepackage[english]{babel}
\usepackage{xspace}
\usepackage{epstopdf}
\usepackage[final]{microtype}

\newcommand{\qql}{\textquotedblleft}
\newcommand{\qqr}{\textquotedblright\xspace}

\newcommand{\hc}{\hat{c}}
\newcommand{\hd}{\hat{c}^\dagger}
\newcommand{\hgamma}{\hat{\gamma}}
\newcommand{\hgammad}{\hat{\gamma}^\dagger}

\newcommand{\deriv}[2]{\frac{\mathrm{d}#1}{\mathrm{d}#2}}

\newcommand{\meanv}[1]{\langle #1 \rangle}

\newcommand{\epower}[1]{\mathrm{e}^{#1}}

\graphicspath{{./Figures/}}

\begin{document}
\title{Flat band transport and Josephson effect through a finite-size sawtooth lattice}
\author{Ville A. J. Pyykk\"onen}
\affiliation{Department of Applied Physics, Aalto University School of Science,
FI-00076 Aalto, Finland}

\author{Sebastiano Peotta} 
\affiliation{Department of Applied Physics, Aalto University School of Science, FI-00076 Aalto, Finland}
\affiliation{Computational Physics Laboratory, Physics Unit, Faculty of Engineering and
Natural Sciences, Tampere University, FI-33014 Tampere, Finland}
\affiliation{Helsinki Institute of Physics, FI-00014 University of Helsinki, Finland}

\author{Philipp Fabritius}
\author{Jeffrey Mohan}
\author{Tilman Esslinger}
\affiliation{Department of Physics, ETH Zurich, 8093 Zurich, Switzerland}

\author{P\"aivi T\"orm\"a}
\email{paivi.torma@aalto.fi}
\affiliation{Department of Applied Physics, Aalto University School of Science, FI-00076 Aalto, Finland}

%


\begin{abstract}
We study theoretically the transport through a finite-size sawtooth lattice coupled to two fermionic reservoirs kept in the superfluid state. We focus on the DC Josephson effect and find that the flat band states of the sawtooth lattice can support larger critical current and at higher temperature than the dispersive band states. However, for this to occur the boundary states of the finite-size lattice need to be tuned at resonance with the bulk flat band states by means of additional boundary potentials. We show that transport in a two-terminal configuration can reveal the salient features of the geometric contribution of flat band superconductivity, namely the linear dependence of key quantities, such as the critical current and critical temperature, on the interaction. Our results are based on parameters of a realistic experimental lattice potential, and we discuss the conditions one needs to reach to observe the predicted effects experimentally.    
\end{abstract}

\maketitle

\section{Introduction}
\label{sec:introduction}

A flat band is a Bloch band of a lattice model which is dispersionless, usually as a consequence of destructive quantum interference between alternative hopping paths. This means that all the states in the band are degenerate and localized, while the kinetic energy of noninteracting particles is completely quenched. The massive degeneracy of  flat bands leads to the strongly-correlated regime and new emergent phases as soon as interparticle interactions are switched on~\cite{leykam2018}.
Indeed, at the theoretical level, flat bands have been proposed to host
ferrimagnetic~\cite{lieb1989} and ferromagnetic phases~\cite{mielke1991a,mielke1991b,
tasaki2008, derzhko2015,costa2016}, various topological states~\cite{sun2011, tang2011, neupert2011, zhao2012, jaworowski2015},
Wigner crystallization \cite{wu2007}, and superconductivity, the main focus of this work.
Lattice models with flat bands can be realized experimentally with
optical lattices for ultracold atoms~\cite{jo2012,taie2015,Leung:2020},
photonic lattices \cite{gersen2005, mukherjee2015, vincencio2015, nguyen2018},
polaritons \cite{jacqmin2014}, but also in van der Waals materials \cite{cao2018,marchenko2018, li2018,po2018}
and artificial electronic systems \cite{drost2017, slot2017, huda2020}.

Flat bands potentially enable high temperature
superconductivity, up to room temperature \cite{kopnin2011,heikkila2011}.
It has been shown theoretically that superconductivity occurs in flat bands
in the presence of attractive interactions only if the band has a
non-trivial quantum metric \cite{peotta2015,julku2016,liang2017}, which is an invariant of the band structure related to the Berry curvature.
More specifically, it has been found that in the flat band limit the geometric contribution of superconductivity~\cite{peotta2015} dominates and the superfluid weight is \textit{linearly proportional}
to both the coupling constant of the attractive interparticle interaction
and the integral of the quantum metric over the first Brillouin zone. A nonzero superfluid weight is the defining property of superfluid/superconducting states.

Predictions of flat band superconductivity are supported by the recent remarkable discovery of 
superconductivity
in magic-angle twisted bilayer graphene~\cite{cao2018}, which is believed to be a direct consequence of the nearly flat bands that occur in the band structure at a specific (\qql magic\qqr) twist angle between the two graphene layers. Indeed, it has been shown in three distinct theoretical studies~\cite{julku2020, hu2019, xie2020, classen2020}
that the geometric contribution to the
superfluid weight, that is the part of the superfluid weight proportional to the band quantum metric, is important in magic-angle twisted bilayer graphene. However, this evidence is only indirect since twisted bilayer graphene is a complex material and some important questions are still open, such as the origin of the attractive interaction responsible for superconductivity~\cite{wu2018, bernevig2020}. 
Moreover, the degree of control on the material properties achieved on this material, for instance by tuning the twist angle, is outstanding, but still too limited to provide direct evidence of the effects associated with the flat band quantum metric. An experiment taking advantage of the degree of control available in ultracold gas experiments~\cite{torma2014} is a highly promising platform for investigating the role played by quantum geometry in a flat band superfluid, as for instance the interaction can be tuned to verify the expected linear dependence of the geometric contribution of superconductivity on interaction.

In order to study flat band superconductivity in the ultracold gas context, we propose and simulate here an atomtronic two-terminal transport setup in which a finite-size sawtooth lattice is placed in contact at the two ends with two superfluid fermionic reservoirs. Two-terminal transport setups are commonly employed to probe solid state mesoscopic systems~\cite{datta1995} and more recently transport experiments of this kind have been realized also using ultracold fermionic lithium atoms~\cite{Brantut2012,krinner2015,husmann2015,lebrat2018,krinner2017,lebrat2018}. In this way the quantization of conductance has been observed in neutral matter for the first time~\cite{krinner2015}. The main goal of atomtronics \cite{Amico2017,Amico2020} is to achieve a high degree of control on ultracold atom transport and ultimately realize complex working devices as in electronics.

Ultracold atoms offer new possibilities with no counterpart in electronics. Two are particularly important for the present work: first, the control on inter-atomic interactions by means of Feshbach resonances and, second, the ability to flexibly engineer complex lattice potentials in the region between the reservoirs (the \qql scattering region\qqr in the Landauer picture of transport) using for instance a digital micromirror device (DMD)~\cite{lebrat2018}. By tuning the inter-atomic interaction strength it should be possible to show that the superfluid weight is \textit{linearly proportional to the interaction}, which is the fingerprint of the effect of the quantum metric on the superfluid properties. This is much harder to do in the solid state context since the interaction strength can be tuned only to a limited extent. Concerning the lattice potentials, digital micromirror devices or more traditional optical lattices can be used to realize complex potentials that implement lattice models with flat bands in the tight-binding limit. A viable optical lattice scheme has been proposed for instance in the case of the sawtooth ladder~\cite{Huber:2010,zhang2015}, which is the lattice model considered in this work.

In a two-terminal setup, it is only possible to insert a finite portion of an infinite lattice model in between the two reservoirs. This creates a non-trivial problem in the case of lattice models with flat bands, which has no counterpart for dispersive bands. The problem is due to the localized nature of the states which compose a flat band: when the infinite lattice is truncated, the flat band states away from the ends of the finite system (the \qql bulk\qqr states) are essentially unaffected precisely because they are strongly localized on a few lattice sites that are not directly connected to the boundary lattice sites by hopping matrix elements \cite{sutherland1986}. On the other hand, the few (usually two) flat band states localized on lattice sites at the two ends of the finite system are strongly affected by the truncation and in general they are not anymore degenerate with the bulk states. These \qql boundary\qqr states are essential for the current to flow from the reservoirs to the bulk of the finite-size system, thus the loss of degeneracy has usually the effect of dramatically suppressing transport through the flat band, as it is shown in the following in the case of the sawtooth ladder. This is an effect specific to flat bands and it is not observed in the case of dispersive bands whose states have a delocalized, plane wave-like character.
 
A key result of this work is that this problem can be solved by properly tuning the energy of the boundary states. Due to the localized character of the flat band states, this can be easily done by introducing additional potential terms at the two ends, which have the purpose of restoring the degeneracy between bulk and boundary states. We expect this problem to be present in various forms of transport, such as in the steady state when the system is driven out of equilibrium by a chemical potential difference. However, in this work we restrict ourselves to the case of transport at equilibrium between superfluid/superconducting reservoirs, that is to the DC Josephson effect~\cite{josephson1962,meden2019}. This is technically the simplest case to handle because the Josephson effect occurs at equilibrium in the presence of a phase difference between the two reservoirs. Moreover, it is more directly connected to previous results on the superfluid weight in infinite lattices~\cite{peotta2015, julku2016, liang2017}. 
Indeed, the superfluid weight is the coefficient of  proportionality between the superfluid current and the phase gradient in the bulk~\cite{scalapino1992,scalapino1993}, while the Josephson critical current is the coefficient of proportionality between the Josephson current and the phase bias, in the limit of small bias. Thus, it is evident that the large superfluid weight in a flat band with nonzero quantum metric, as predicted in previous works~\cite{peotta2015, liang2017, tovmasyan2018},
translates into an expected large Josephson critical current in  a two-terminal setup. This expectation is fully confirmed in this work, under the condition that the degeneracy between boundary and bulk states is restored, as briefly explain above and with extensive details in the following. The Josephson effect, without any connection to flat band physics, has been widely studied in ultracold gas experiments~\cite{Cataliotti2001,albiez2005,Levy2007,valtolina_josephson_2015,Betz2011,Ryu2013,Spagnolli2017,Burchianti2018}, and most recently observed in superfluid Fermi gas \cite{Luick2020, Kwon2020}.

The paper is organized as follows. Section~\ref{sec:model} presents the model and 
in Section~\ref{sec:method} the method used to compute the Josephson current is described in detail.
In Section~\ref{sec:Josephson_current}, we first discuss results in the case when interactions are not present within the finite-size sawtooth lattice,
and then show how the Josephson current is modified by interactions.
In Section~\ref{sec:boundary} we include the boundary potentials that allow to restore the degeneracy between bulk and boundary states, as discussed above, and show that the Josephson current through the flat band is strongly enhanced at resonance.
In Section~\ref{sec:finite_temp} we include the effect of a finite
temperature and estimate the superconducting critical temperatures below which a nonzero Josephson critical current can be observed, both in the flat and dispersive band case. In Section~\ref{sec:experimental_realization} a possible experimental realization with ultracold gases is proposed based on an optical potential that can be used to implement the sawtooth lattice.
Finally, in Section~\ref{sec:discussion} we discuss and summarize the results.

\section{Methods}
\subsection{Tight-binding model Hamiltonian}
\label{sec:model}

\begin{figure}
    \centering
    \includegraphics[width=\columnwidth]{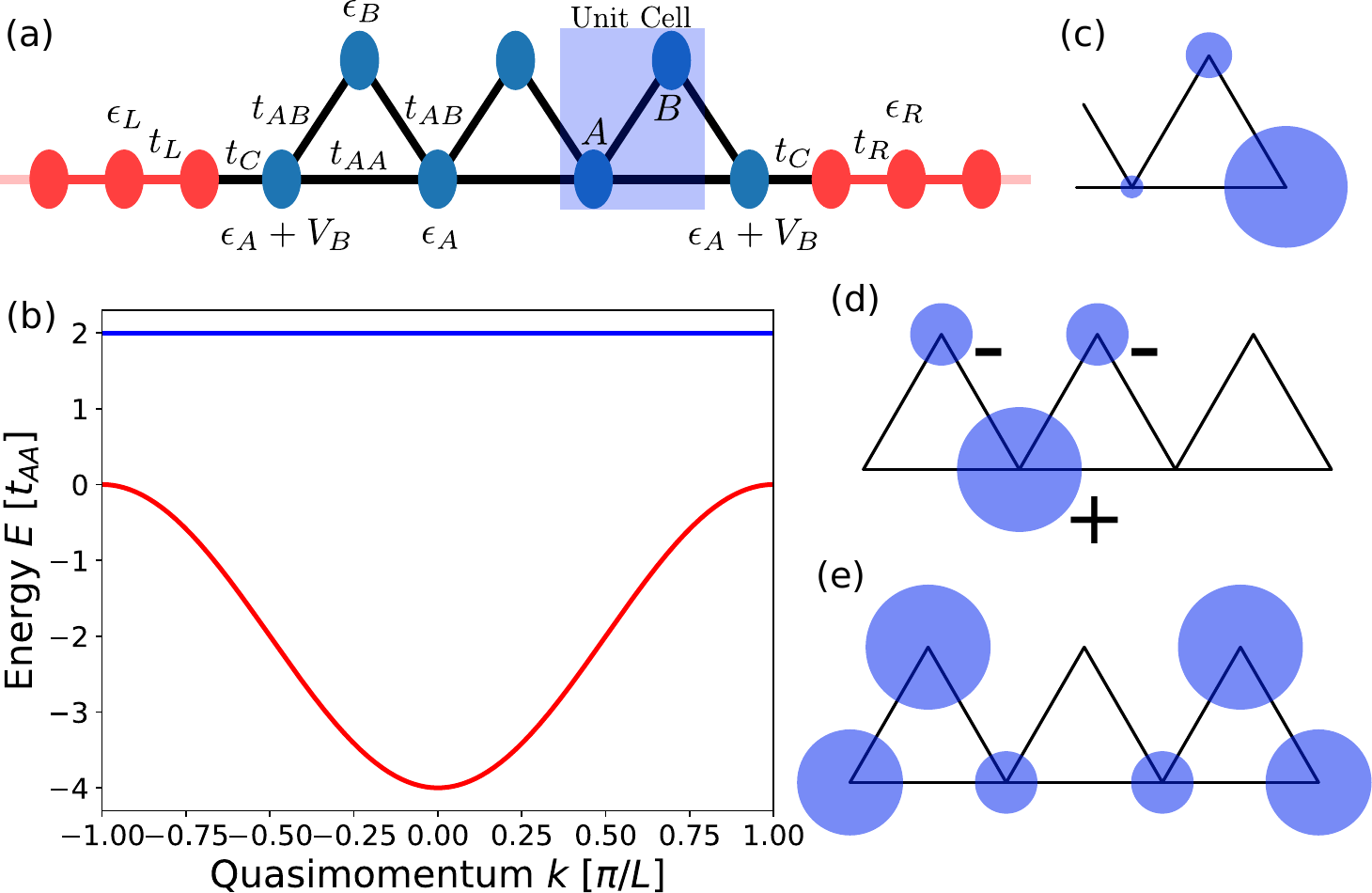}
    \caption{(a) Description of the two-terminal setup tight-binding model as a graph where the sites represent on-site energies and the edges the non-zero hopping amplitudes between sites.
    The figure introduces the notation and shows the structure of the sawtooth ladder and the leads. The leads (red
    sites) are semi-infinite. (b) The band structure
    of an infinite sawtooth ladder with parameters $\epsilon_A=\epsilon_B = 0, t_{AB} = \sqrt{2}t_{AA}.$ Figures (c)-(e) show examples of different eigenstate types of the finite-size sawtooth ladder. The sizes of the circles at the lattice sites indicate the wavefunction amplitude squared, that is, the relative probabilities. (c) An edge state. (d) A flat band state. The signs $\pm$ denotes the respective phases of the wavefunction at the sites. (e) A dispersive band state.}
    \label{fig:flat_band_system}
\end{figure}

We consider a two-terminal setup comprising a finite portion of a sawtooth ladder
and two leads (the reservoirs) described by the
following non-interacting tight-binding Hamiltonian 
\begin{equation}
    \hat{H}_0 = \hat{H}_{\mathrm{sawtooth}} + \hat{H}_{L}+\hat{H}_{R} +
    \hat{H}_{\mathrm{contact}}~,
    \label{normaltbH}
\end{equation}
where $\hat{H}_{\mathrm{sawtooth}}$ describes the sawtooth lattice,
$\hat{H}_{L},\hat{H}_{R}$ describes the two leads, left and right respectively, 
and $\hat{H}_{\mathrm{contact}}$ describes the contact between leads and the sawtooth ladder.
The tight-binding model is presented in Figure \ref{fig:flat_band_system} (a), where the notation employed here is also introduced.

The sawtooth ladder is a 1D lattice with two sites per unit cell, called $A$ and $B$~\cite{Huber:2010}.
The $A$ sites are connected to the nearest neighbor $A$ sites by 
the hopping amplitude $t_{AA}$ and to the nearest-neighbor $B$ sites
by $t_{AB}$.
The $B$ sites are not connected to each other by direct hopping amplitudes.
At each site there is an on-site energy term $\epsilon_{A}$, $\epsilon_B$ for $A$ and $B$ sites, respectively.
The sawtooth ladder Hamiltonian is 
\begin{equation}
    \begin{split}
        \hat{H}_{\mathrm{sawtooth}} &= \sum_{i=1}^{N_{\rm c}}\sum_\sigma\Big[\big(
        \epsilon_A\hat{c}_{A,i\sigma}^\dagger\hat{c}_{A,i\sigma} + \epsilon_B
        \hat{c}_{B,i\sigma}^\dagger\hat{c}_{B,i\sigma}\big) \\& -
        \big(t_{AB}\hat{c}_{B,i\sigma}^\dagger\hat{c}_{A,i\sigma} +
        t_{AB}\hat{c}_{A,i+1\sigma}^\dagger\hat{c}_{B,i\sigma}\\ 
        &+ t_{A}\hat{c}_{A,i+1\sigma}^\dagger\hat{c}_{A,i\sigma} + \mathrm{H.c.}\big)\Big] \\ &+ \sum_{\sigma}
        \epsilon_A\hat{c}_{A,N+1\sigma}^\dagger\hat{c}_{A,N+1\sigma},
    \end{split}
\end{equation}
where $N_{\rm c}$ is the number of unit cells and 
$\hat{c}_{\alpha,i\sigma}$, $\hat{c}^\dagger_{\alpha,i\sigma}$ annihilates and creates, respectively, a particle at the site $\alpha = A/B$ of unit cell $i$ with spin $\sigma = \{\uparrow,\downarrow\}$.

The sawtooth ladder has two sites per unit cell so it contains two bands, one of which is flat if the hopping amplitudes satisfy the condition  $t_{AB} = \sqrt{2t_{AA}^2 + (\epsilon_B-\epsilon_A)t_{AA}}$, and the other is dispersive. We consider here the case $\epsilon_A=\epsilon_B$ and $t_{AB} =\sqrt{2}t_{AA}$, which gives the band structure in Figure~\ref{fig:flat_band_system} (b). The Hamiltonian of the sawtooth ladder with $N_{\rm c}$ unit cells has $2N_{\rm c}+1$ eigenstates, $N_{\rm c}$ of which are related to the
dispersive band, $N_{\rm c}-1$ are flat band states and $2$ are edge or boundary states. The localized flat band
states, where contributions outside a V shaped region vanish due to destructive
interference, are shown in Figure \ref{fig:flat_band_system} (d). 
Notice that the localized states shown in Figure \ref{fig:flat_band_system} (d) are not orthogonal to each other.
Nevertheless, the states span the flat band subspace, and a proper orthonormal basis 
can be constructed comprising very similar states, which however possess exponentially 
decaying tails~\cite{Huber:2010}.
An example of a state of the dispersive band is shown in Figure \ref{fig:flat_band_system}
(e). The dispersive band states are spread over the whole system. An example of
edge state is shown in Figure \ref{fig:flat_band_system} (c). It has major
contribution at one edge and decays exponentially as function of distance 
from the edge.

The leads are modeled as simple chains with local on-site energy $\epsilon_L$, $\epsilon_R$ where $L,R$ refer to left and right leads,
respectively.
The lead bandwidth is controlled by the hopping amplitude between sites
$t_{L/R}$.
The leads are connected to the sawtooth ladder at the edge $A$ sites
by contact hopping amplitudes $t_C.$
The leads are modeled by the Hamiltonians
\begin{equation}
    \begin{split}
    \hat{H}_{L/R}
    = \sum_{i,\sigma} &\Big[\epsilon_{L/R} \hat{c}_{L/R,i\sigma}^\dagger
    \hat{c}_{L/R,i\sigma} \\ 
    &-t_{L/R}\big(\hat{c}_{L/R,i+1\sigma}^\dagger\hat{c}_{L/R,i\sigma} + \mathrm{H.c.}\big)\Big]\,,
    \end{split}
\end{equation}
where $\hat{c}_{L/R,i\sigma}^\dagger, \hat{c}_{L/R,i\sigma}$ are the creation and annihilation operators for the leads. The unit cell index $i = \{1,2,\dots\}$ in the lead operators increases from
the edge. 
Finally, the contact Hamiltonian is 
\begin{equation}
\begin{split}
    \hat{H}_{\mathrm{contact}}
    =  \sum_{\sigma}&\Big[-t_C\big(\hat{c}_{L,1\sigma}^\dagger\hat{c}_{A,1\sigma}  + \hat{c}_{R,1\sigma}^\dagger\hat{c}_{A,N+1\sigma}  + \mathrm{H.c.}\big) \\
    &+ V_B \big(\hat{c}_{A,1\sigma}^\dagger \hat{c}_{A,1\sigma}
    + \hat{c}_{A,N+1\sigma}^\dagger \hat{c}_{A,N+1\sigma}  \big)\Big]\,,
    \end{split}
\end{equation}
where $t_C$ is the tunneling amplitude between lead and the sawtooh lattice and $V_B$ is a boundary potential introduced to tune the edge states energy and restore degeneracy as discussed in Section~\ref{sec:introduction}.

\subsection{Interacting Hamiltonian and self-consistent mean field method}
\label{sec:method}

The full many-body grand canonical Hamiltonian takes the form
\begin{equation}
    \hat{H} =\hat{H}_0 +\hat{H}_{\mathrm{int}} - \mu \hat{N}~,
    \label{eq:int_ham}
\end{equation}
where the interaction term $\hat{H}_{\mathrm{int}}$ is the Hubbard interaction
\begin{equation}
    \hat{H}_{\mathrm{int}} = -\sum_{i} U_\alpha
    \hat{c}_{\alpha, i\uparrow}^\dagger\hat{c}_{\alpha,i\uparrow}\hat{c}_{\alpha,i\downarrow}^\dagger
    \hat{c}_{\alpha, i\downarrow}\,,
\end{equation}
$\mu$ is the chemical potential and $\hat{N} = \sum_{\alpha,i\sigma} \hat{c}^\dagger_{\alpha,i\sigma}\hat{c}_{\alpha,i\sigma}$ is the number operator. 
Here the index $\alpha = \{L,R,A,B\}$ labels the various parts of the system,
$U_\alpha \geq 0$ (attractive interaction) is the interaction strength,
and $i$ goes over the unit cells of the tight-binding model.
To solve the many-body problem, we use  
the Bardeen-Cooper-Schrieffer (BCS) mean-field theory in the form of the Bogoliubov-Valatin
canonical transformation. 
In the case of flat bands, the mean-field theory is expected to perform
well for interaction strengths $U_{\alpha}$ up to the band gap between the flat band and its nearest neighbouring band~\cite{tovmasyan2016}. The band gap between the flat band and the dispersive band in the sawtooth lattice is $2t_{AA}.$ 
In the mean field approximation, we
approximate the Hubbard term as follows, up to a constant 
\begin{equation}
\begin{split}
    &U_\alpha\hat{c}_{\alpha, i\uparrow}^\dagger\hat{c}_{\alpha,i\uparrow}\hat{c}_{\alpha,i\downarrow}^\dagger
    \hat{c}_{\alpha, i\downarrow}\\
    &\simeq U_\alpha \bigg(
    \langle\hat{c}_{\alpha, i\uparrow}^\dagger\hat{c}_{\alpha,i\uparrow}\rangle
    \hat{c}_{\alpha,i\downarrow}^\dagger \hat{c}_{\alpha,i\downarrow} + 
    \langle\hat{c}_{\alpha,i\downarrow}^\dagger \hat{c}_{\alpha,i\downarrow}
    \rangle\hat{c}_{\alpha,i\uparrow}^\dagger\hat{c}_{\alpha,i\uparrow} \\ &+
    \langle\hat{c}_{\alpha,i\uparrow}^\dagger \hat{c}_{\alpha,i\downarrow}^\dagger
    \rangle\hat{c}_{\alpha,i\downarrow}\hat{c}_{\alpha,i\uparrow} +
    \langle\hat{c}_{\alpha,i\downarrow}\hat{c}_{\alpha,i\uparrow} 
    \rangle\hat{c}_{\alpha,i\uparrow}^\dagger \hat{c}_{\alpha,i\downarrow}^\dagger\bigg)~.
     \end{split}
\end{equation}
By utilizing the canonical commutation relations, we write
the system Hamiltonian in Nambu form with the vectors
\begin{equation}
    \hat{d}_{\alpha,i} = \begin{pmatrix} \hat{c}_{\alpha,i\uparrow} \\ \hat{c}_{\alpha,j\downarrow}^\dagger \end{pmatrix},
    \quad
    \hat{d}_{\alpha,i}^\dagger = \begin{pmatrix} \hat{c}_{\alpha,i\uparrow}^\dagger & \hat{c}_{\alpha,j\downarrow} \end{pmatrix}~.
\end{equation}
In this basis, we have up to a constant
\begin{equation}
    \hat{H} \simeq \sum_{\alpha i,j\beta}\hat{d}_{\alpha,i}^\dagger \mathcal{H}_{\mathrm{BdG},\alpha i,\beta j} \hat{d}_{\beta,j} ~,
\end{equation}
where we have defined the Bogoliubov-de Gennes (BdG) Hamiltonian $\mathcal{H}_{\mathrm{BdG}}$ as 
\begin{equation}
\begin{split}
    \mkern-4mu&\mathcal{H}_{\mathrm{BdG},\alpha i, \beta j}\\
    \mkern-4mu =& 
    \begin{pmatrix}
        T_{\alpha i,\beta j} + V_{\alpha i\downarrow}\delta_{\alpha i,\beta j} & \Delta_{\alpha i} \delta_{ \alpha i, \beta j} \\
        \Delta_{\alpha i}^*\delta_{\alpha i,\beta j} & -T_{\alpha i,  \beta j} - V_{\alpha i\uparrow}\delta_{\alpha i,\beta j}\\
    \end{pmatrix}.
    \end{split}
\end{equation}
Here 
\begin{equation}
    T_{\alpha i, \beta j} = (\epsilon_{\alpha,i}-\mu)\delta_{\alpha i,\beta j} - t_{\alpha i , \beta j}
    \label{eq:single_particle}
\end{equation}
includes all of the single-particle terms of the grand canonical Hamiltonian \eqref{eq:int_ham},
the superconducting order parameter $\Delta_{\alpha i}$ is given by 
the gap equation
\begin{equation}
\label{eq:Delta}
    \Delta_{\alpha i} = -U_\alpha \meanv{\hat{c}_{\alpha, i\downarrow}\hat{c}_{\alpha, i\uparrow}}
\end{equation}
and the Hartree potential 
$V_{\alpha i\sigma}$ is given by 
\begin{equation}
    V_{\alpha  i\uparrow/\downarrow} = -U_\alpha \meanv{\hat{c}_{\alpha, i\downarrow/\uparrow}^\dagger \hat{c}_{\alpha, i\downarrow/\uparrow}}~.
\end{equation}
In this work, we assume time-reversal symmetry which implies
$V_{\alpha i \uparrow} = V_{\alpha i \downarrow}  \equiv V_{\alpha i}$.

The mean-field BdG Hamiltonian is diagonalized as $\mathcal{H}_{\rm BdG} = SDS^\dagger$, where $D$ is a diagonal
matrix containing the eigenvalues of $\mathcal{H}_{\rm BdG}$ and $S$ is the unitary matrix comprising
the corresponding eigenvectors as columns, in the respective order.
In the diagonalized basis, the Hamiltonian becomes
\begin{equation}
    \hat{H} = \sum_{n\sigma} E_{n} \hat{\gamma}_{n\sigma}^\dagger \hat{\gamma}_{n\sigma}
\end{equation}
where $E_{n}$ are the eigenvalues contained in $D$ and
the quasiparticle operators $\gamma_{n\sigma}$ are defined by the following Bogoliubov-Valatin transformations
\begin{equation}
    \begin{split}
        \hc_{\alpha,i\uparrow} &= \sum_n\left(u_{\alpha i,n} \hgamma_{n\uparrow} + v^*_{\alpha i,n} \hgammad_{n\downarrow}\right)\\
        \hd_{\alpha,i\downarrow} &= \sum_n\left(v_{\alpha i,n} \hgammad_{n\uparrow} - u_{\alpha i,n}^* \hgammad_{n\downarrow}\right) ~,
    \end{split}
\end{equation}
with coefficients $u_{\alpha i,n}$, $v_{\alpha i,n}$ given by $u_{\alpha i,n} = S_{2(\alpha i)-1,n}$, $v_{\alpha i,n} = S_{2(\alpha i),n}$ related to positive eigenenergies $E_n$, where $\alpha i=\{1,2,\dots\}$ denotes the index corresponding to the site of unit cell $i$ in sublattice $\alpha$ in the indexing of the single-particle Hamiltonian matrix \eqref{eq:single_particle}.  
In the mean-field approximation, the quasiparticles are non-interacting and thus obey the Fermi-Dirac statistics.

Using the Bogoliubov-Valatin transformation in Eq.~\eqref{eq:Delta},
one obtains the gap equation
\begin{equation}
    \Delta_{\alpha i} = U_\alpha \sum_n u_{\alpha i,n}v_{\alpha i,n}^* \tanh\left(\frac{\beta E_n}{2}\right)~,
    \label{gap_eq}
\end{equation}
where $E_n$ is the energy of the respective BdG Hamiltonian eigenstate $n$
and $\beta=1/T$ is the inverse temperature (in our units Boltzmann's constant $k_B=1$)
and similarly the Hartree potential is given by
\begin{equation}
    V_{\alpha i} = -\sum_{n}\left( \frac{U_\alpha|u_{\alpha i,n}|^2}{\exp(\beta E_n)+1} + \frac{U_\alpha|v_{\alpha i,n}|^2}{\exp(-\beta E_n) +1 }\right)~.
    \label{Hartree_eq}
\end{equation}

In the leads, the order parameters $\Delta_{L/R,i}$ are set to a constant $\Delta_{L}$,
while the Hartree potential $V_{L/R,i}$ are put to zero.
The finite Josephson current is the result of having the lead
superconducting order parameters equal in amplitude but with different phases.

The order parameters $\Delta_{A/B,i}$ and the Hartree potentials $V_{A/B,i}$ are
calculated self-consistently.
In practice, the values for the self-consistent parameters are determined by the following
iterative algorithm. 
\begin{enumerate}
    \item Give initial guesses for the self-consistent parameters $\Delta_{\alpha i}$ and $V_{\alpha i}$.
    \item Diagonalize the BdG Hamiltonian $\mathcal{H}_{\mathrm{BdG}}$.
    \item Update the self-consistent parameters $\Delta_{\alpha i}$ and $V_{\alpha i}$ by using Eqs.~\eqref{gap_eq}-\eqref{Hartree_eq} and the eigenvalues and eigenvectors of the BdG Hamiltonian obtained in Step 2.
    \item If the difference between the updated and the pre-update parameters is less  than the wanted accuracy (in an appropriate norm), end the procedure. Otherwise, go back to Step 2.
\end{enumerate}
Oftentimes this naive iterative procedure does not converge, leading to an 
oscillating solution. This issue can be solved by adopting a mixing algorithm,
where self-consistent parameters of previous iterations are mixed at the Step 3 with the new values to give the update.
For small enough mixing of the new iterate, the algorithm converges \cite{yang2008}.
However, the convergence can be arbitrarily slow in general.
To boost the convergence, we use two algorithms in combination with the simple mixing:
the Broyden's method \cite{broyden1965}, which is a pseudo-Newton iteration,
and the Anderson-Pulay mixing \cite{anderson1965,pulay1982}.

The current operator for the system is obtained by
the continuity equation and Heisenberg equation of motion for the particle number operator $\hat{n}_{\alpha,i}  = \sum_\sigma\hat{c}_{\alpha,i\sigma}^\dagger \hat{c}_{\alpha,i\sigma}$, which leads to
\begin{equation}
    \deriv{\hat{n}_{\alpha,i}}{t} = \sum_{\beta j} I_{\alpha i, \beta j} + K_{\alpha,i}\,,
\end{equation}
where 
\begin{gather}
    I_{\alpha i,\beta j}
     =- \frac{i}{\hbar}
     \sum_{\sigma} t_{\alpha i,\beta j}\left(\meanv{\hat{c}_{\alpha,i\sigma}^\dagger
     \hat{c}_{\beta,j\sigma}} + \meanv{\hat{c}_{\beta,j\sigma}^\dagger
     \hat{c}_{\alpha,i\sigma}}\right)\\
     K_{\alpha,i} = \frac{i}{\hbar}
     \left(
     \Delta_{\alpha i}^*\meanv{\hat{c}_{\alpha,i\downarrow}\hat{c}_{\alpha,i\uparrow}}
     -\Delta_{\alpha i} \meanv{\hat{c}_{\alpha,i\uparrow}^\dagger\hat{c}^\dagger_{\alpha,i\downarrow}}
     \right)\,,
\end{gather}
where $I_{\alpha i,\beta j}$ is the current from site $\beta j$ to site $\alpha i$,
and $K_{\alpha,i}$ is a source term.
When the superconducting order parameter is calculated self-consistently, $K_{\alpha,i}$ vanishes and particle conservation is ensured.

The leads are truncated to finite length so that the
system is finite and closed, as in Ref. \cite{black-schaffer2008}.
In the leads, however, the superconducting order parameter is constant and not calculated self-consistently, therefore $K_{\alpha,i}$ is finite and particle number is not conserved. This makes it possible to have  a finite equilibrium current in a
closed system. The Josephson current is obtained by evaluating the current expectation value
\begin{equation}
    I_{\alpha i, \beta j}
     = \frac{2}{\hbar}
     \sum_{\sigma} 
     t_{\alpha i,\beta j}\mathrm{Im}\left(\meanv{\hat{c}_{\alpha,i\sigma}^\dagger
    \hat{c}_{\beta,j\sigma}}\right) ~,
\end{equation}
which can be written in terms of the Bogoliubov-Valatin transformation parameters using
\begin{equation}
    \meanv{\hat{c}_{\alpha,i\sigma}^\dagger\hat{c}_{\beta,j\sigma}}
    =\sum_n \left( \frac{u_{\alpha i,n}^*u_{\beta j,n}}{\epower{ \beta E_n}+1}
    +\frac{v_{\alpha i,n}v_{\beta j,n}^*}{\epower{-\beta E_n}+1}\right)~.
\end{equation}

\section{Results}
\subsection{Josephson current through a sawtooth ladder}
\label{sec:Josephson_current}

We first present results for the Josephson current when $U_{A/B}=0$,
which we refer to as the non-interacting case, and then in
the interacting case $U_{A/B}\neq 0.$ Note that the leads are always assumed to be superconducting; the interactions necessary to induce superconductivity in the leads are not explicitly considered but are implicit in the constant superconducting order parameter of the leads. Thus non-interacting/interacting refers to only to the sawtooth lattice in the transport channel.  
The non-interacting case is presented here because it is a useful  reference for understanding the interacting case.
Indeed, it illustrates how the various states of the finite-size sawtooth ladder affect the Josephson current.

We consider leads of $20$ sites and
a sawtooth ladder with $N_{\rm c} =3$ unit cells.
In order to make the results relevant for ultracold gas systems,
the parameters for the sawtooth ladder tight-binding model are extracted
from a realistic potential that can be realized experimentally with a digital micromirror device for instance, shown in Figure \ref{fig:sawtooth_potential}. See Section~\ref{sec:experimental_realization} for more details on a possible experimental realization.
The lead parameters $t_{L/R} = 30\,{\rm kHz}$ are set so that the 
lead bandwidth is wide, that is, large in comparison to the other energy scales in the system.
In Figure~\ref{fig:sawtooth_potential} we collect all the sawtooth lattice parameters that are always fixed for the results presented here. The energy unit is the hertz (Hz) since we set $\hbar=1$.
This scale is appropriate for ultracold gas systems since 1 nK in temperature
corresponds to 20.84 Hz and the usual temperatures are of order
60-70 nK \cite{krinner2017,lebrat2018}. The parameters not specified in Figure~\ref{fig:sawtooth_potential} are varied in different analyses
and provided separately. We always put a constant order parameter $\Delta_L$ and zero Hartree potential $V_{L} = 0$ in the leads, moreover in the non-interacting case the
order parameter $\Delta_{A/B}$ and Hartree potential $V_{A/B}$ ladder vanish  within the sawtooth ladder.  The chemical potential $\mu$ is tuned 
in order to control the filling of the states within the sawtooth ladder.
For each value of the chemical potential, we vary the superconducting phase difference between the leads from $0$ to $2\pi$ and determine the maximal current, known as the critical Josephson current.

\begin{figure*}
    \centering
    \includegraphics[width=\textwidth]{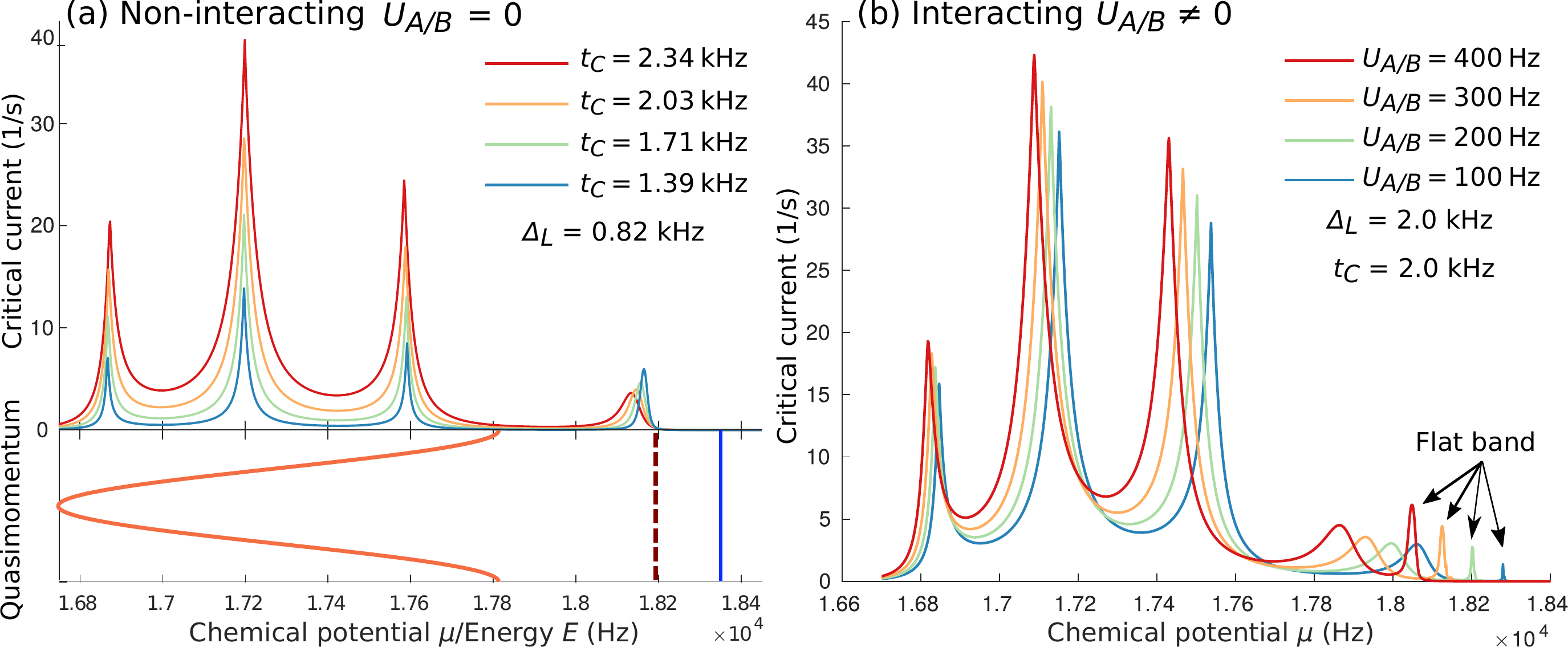}
    \caption{
        (a)
        Critical Josephson current vs. chemical potential $\mu$ for a sawtooth ladder with $N_{\rm c} = 3$ unit cells without interactions
     ($U_{A/B}=\Delta_{A/B,i} =V_{A/B,i}=0$). In the lower panel, the band structure of an infinite sawtooth lattice, with the edge state energy indicated as a dashed line, is shown for comparison.  The
    dispersive band states are responsible for the first three peaks from the left, and are situated within the dispersive energy band of the infinite sawtooth ladder. There are
    $N_{\rm c}$ dispersive band states in a finite-size sawtooth ladder with $N_{\rm c}$ unit cells. The fourth peak is caused by the two almost degenerate
    edge states. The flat band states (marked with blue in the dispersion relation) are not visible since non-interacting particles are localized in these states.
    (b) Critical Josephson current vs chemical potential through the interacting sawtooth ladder for various interaction strengths $U_{A/B}$. We observe the peaks of the same origin as in the non-interacting case of Figure (a) and additionally a peak corresponding to the two bulk flat band states (see Figure~\ref{fig:flat_band_system} (d)).}
    \label{fig:interacting_sawtooth}
\end{figure*}

The non-interacting results at zero temperature are shown in 
Figure \ref{fig:interacting_sawtooth} (a). The critical Josephson current, that is, the maximal
Josephson current with respect to the phase variation, is shown as a function of the chemical
potential.
As a comparison, the band structure of the infinite sawtooth ladder is shown
for values of the energy corresponding to the chemical potential.
It is seen that there are critical Josephson current
peaks corresponding to the eigenstates of the finite size sawtooth ladder.
We observe three peaks corresponding to the dispersive band and a peak corresponding
to the two almost degenerate edge states. There is no peak corresponding to the flat band
due to the fact that non-interacting particles are localized in the flat band states.
The width of the peaks is related to $t_C$.
We see that the positions of the peaks closely matches the band structure.

Next, we include interactions in the finite-size sawtooth ladder as well ($U_{A} = U_{B} \neq 0$) and compute self-consistently the superconducting order parameter $\Delta_{A/B,i}$ and the Hartree potential $V_{A/B,i}$ as discussed in Section~\ref{sec:method}.
The critical Josephson current as a function of the chemical potential is shown in Figure
\ref{fig:interacting_sawtooth} (b). A new critical Josephson current peak is observed, which corresponds to the flat band states.
This agrees with the theoretical expectation based on the study of infinite lattice systems,
since non-interacting particles are localized in flat bands but a finite interaction in geometrically
non-trivial flat bands makes particles non-localized \cite{peotta2015, julku2016, liang2017, torma2018}.
Furthermore, due to interactions, the current through all the states increases
and the peaks are shifted to lower chemical potential
due to the Hartree potential.
The flat band state peak critical Josephson current is found to be linearly dependent
on the interaction strength $U_{A/B}$ as can be expected from the linear dependence of the
superfluid weight on the interaction strength in the case of an infinite lattice \cite{peotta2015}.

We observe that the flat band state current peak is
quite small in comparison to the current associated to dispersive band states
even with interaction strength
of the order of the band gap between the flat band and the dispersive band.

\subsection{Boundary potential}
\label{sec:boundary}

\begin{figure*}
    \centering
    \includegraphics[width=\textwidth]{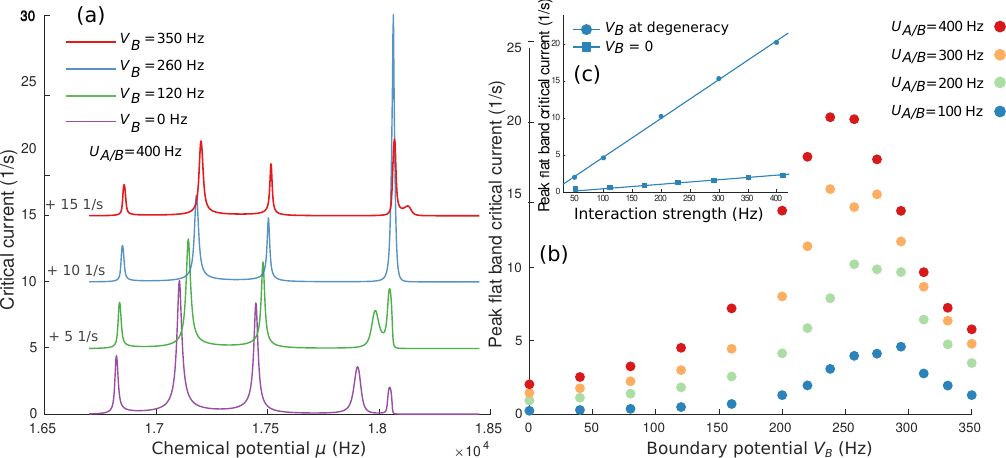}
    \caption{
    (a) Critical Josephson current vs. chemical potential $\mu$ for different boundary
    potentials $V_B$. For clarity, the curves have been shifted up as indicated. 
    When $V_B = 260$ Hz (blue curve) the flat band current is strongly enhanced. Here, $\Delta_L =$ 2.0 kHz and $t_C = $ 1.0 kHz.
    (b) The peak critical Josephson current at the flat band versus boundary
    potential $V_B$ for different interaction strengths $U_{A/B}$.
    It is seen that the current is maximal around a $U_{A/B}$-dependent value of $V_B$ and beyond that the current begins to decrease.
    This $V_B$ value, called the degeneracy value, corresponds to the case in which the bulk flat band states
    and the edge states become degenerate in energy.
    (c) The dependence of the flat band peak critical Josephson current on the interaction strength $U_{A/B}$ 
    both when $V_B$ is tuned at the resonance value and at $V_B=0$. The dependence is found to be linear in both cases but the current is an order of magnitude higher at the degeneracy value of $V_B$. }
    \label{fig:boundary_potential}
\end{figure*}

The observed flat band state peak critical 
current in Figure~\ref{fig:interacting_sawtooth} (b) is small in comparison to the critical currents through the dispersive band states.
As we argued in Section~\ref{sec:introduction}, the reason for this is the loss of degeneracy between bulk and edge states.
In order to restore the degeneracy, we use the boundary potential $V_B$ described in
Section~\ref{sec:model} to tune the energy of the edge states.
The effect of the edge state potential on the current is illustrated in Figure
\ref{fig:boundary_potential}. The parameters are the same as in the previous Section~\ref{sec:Josephson_current} if not specified otherwise.
It is seen that at a certain value of the edge potential $V_B$, the current is increased significantly with respect
to the case $V_B = 0$. We have checked that at this value of the
edge potential, the flat band states and the edge states are degenerate.  
Thus, the hypothesis that the flat band current is significantly increased
when the edge states and flat band states are degenerate seems to be correct.
Importantly, the flat band critical Josephson current is seen to increase \textit{linearly} with the
interaction strength in Figure \ref{fig:boundary_potential} (c).

\subsection{Finite temperature}
\label{sec:finite_temp}

\begin{figure}
    \centering
    \includegraphics[width=\columnwidth]{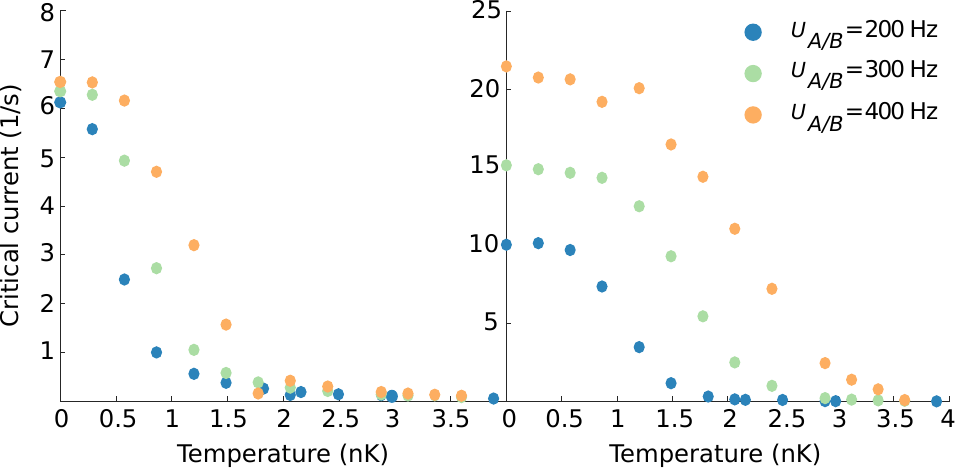}
    \caption{ (a) The peak critical Josephson current dependence on temperature for
    the dispersive band state with the largest current.
    The current vanishes around 1-2 nK, depending
    on the interaction strength $U_{A/B}$.
    (b) The peak critical Josephson current dependence on temperature for
    the flat band states. The boundary potential $V_B$ is set here to the degeneracy value.
    The current vanishes below measurable amplitude around 2-4 nK, depending
    on the interaction strength $U_{A/B}$.
    The dependence of the critical current on the temperature and of the
    critical temperature on the interaction is different for the flat band states
    and the dispersive band states.}
    \label{fig:finite_temp}
\end{figure}

The previous results have been obtained at zero temperature $T=0$. We consider in this section the finite temperature case in order to understand the temperature regions where the experiments are potentially performed and seek for the
different expected temperature dependence of the critical current corresponding to the flat band states and the dispersive band states.
The critical temperatures are determined by finding the temperatures at which the critical current vanishes. We consider the situation with the edge potential $V_B$ tuned to the degeneracy. Otherwise, the parameters are the same as in Figure~\ref{fig:boundary_potential}. The results are shown in 
Figures \ref{fig:finite_temp} (a) and (b). It is seen that the critical temperature of the flat band state is higher than for the dispersive state.
Also, the functional dependency of the critical temperature on the interaction is found to be different for dispersive band states and
the flat band states: for flat bands the dependence is linear but for dispersive band states it is not.
The critical temperature for dispersive band states is found to be in the range 1-2 nK for the considered interaction strengths,
whereas for the flat band it is 2-4 nK.

\section{Experimental realization}
\label{sec:experimental_realization}
\begin{figure}
    \centering
    \includegraphics[width=\columnwidth]{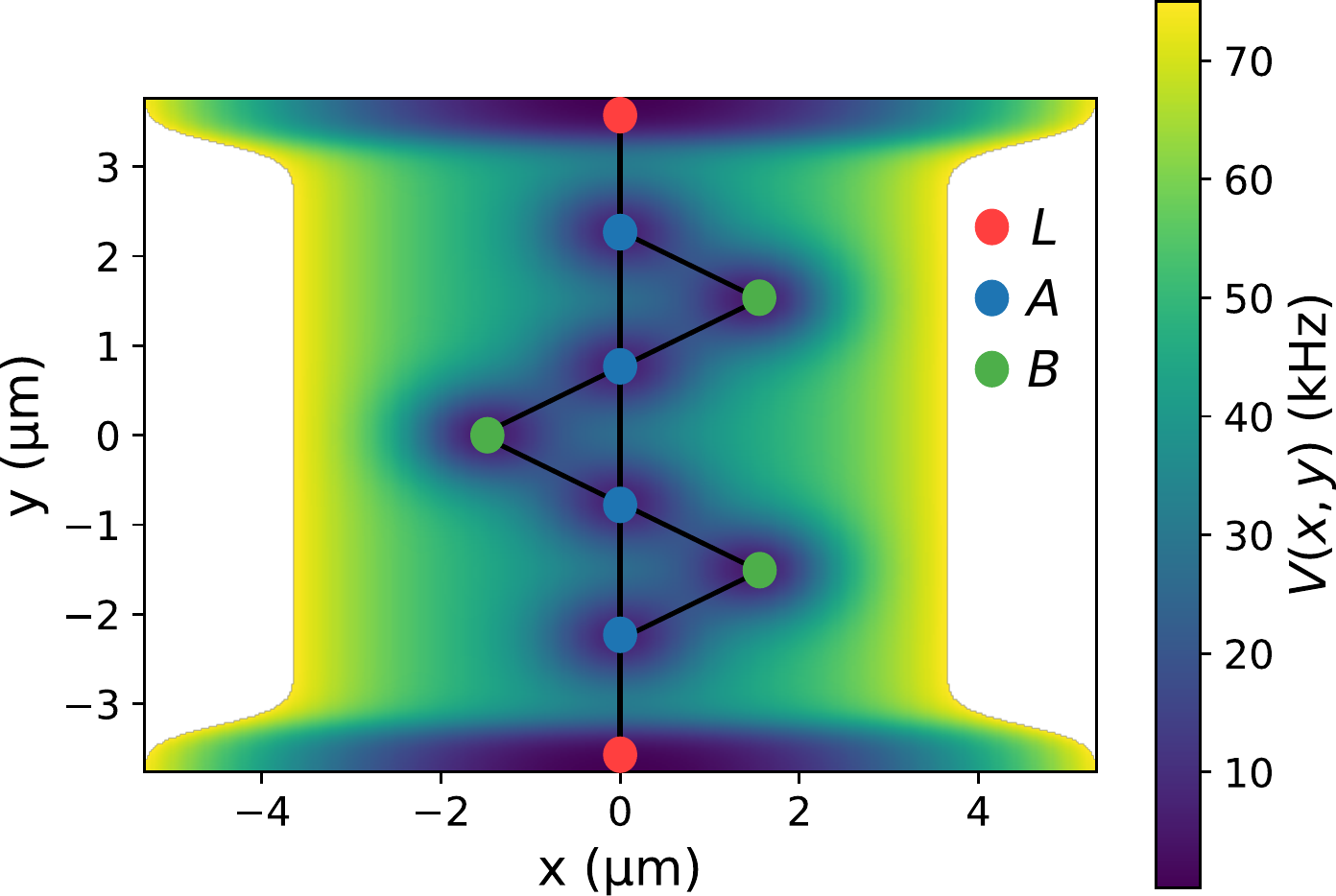}
    \begin{tabular}{c c c c c}
        $\epsilon_A$ & $\epsilon_B$ & $t_{AA}$ & $ t_{AB} $ \\
        \midrule
        17.8 kHz & 17.8 kHz & 271 Hz  & 383 Hz
    \end{tabular}
    \caption{\label{fig:sawtooth_potential} A possible optical potential for implementing the sawtooth ladder in a two-terminal transport experiment with ultracold gases. The corresponding tight-binding model graph is shown on top of the potential. The tight-binding parameters of the sawtooth lattice shown below are extracted from the potential and have been used to produce all of the results presented in this work. }
\end{figure}

Based on the theoretical results shown in Figures \ref{fig:interacting_sawtooth}-\ref{fig:finite_temp}, we now discuss a possible experimental realization in a mesoscopic cold atom transport experiment. Fermionic lithium-6 is an ideal candidate for probing transport in such structures due to their light mass, which leads to high tunneling rates between lattice sites, and tunable interactions from weakly interacting BCS limit to strong interactions (so-called unitarity regime~\cite{torma2014,zwerger2012}). The two leads and the scattering region connecting them can be formed out of a cloud of lithium atoms in a dipole trap by shining two $\mathrm{TEM_{01}}$ beams of blue-detuned, repulsive light. The resulting channel at the intersection of the two beams’ nodal planes (Figure \ref{fig:sawtooth_potential}) permits transport between the two leads which is ballistic in the case described or, if an additional potential like a sawtooth lattice is projected into this region, can have a more complex energy dependence which can be probed by an additional gate beam. 

Transport properties of the scattering region such as the conductance or critical current can be measured by preparing a particle number imbalance between the two reservoirs and measuring the particle number in each reservoir via absorption imaging as a function of time. The critical current can then be probed via coherent Josephson oscillations in the particle imbalance in addition to the normal, dissipative flow as the frequency of these oscillations is directly proportional to the square root of the critical current for small particle and phase imbalances \cite{valtolina_josephson_2015}.
This peculiar dependence of the oscillation frequency on the square root of the critical current is a consequence of the  finite size of the reservoirs, as explained in the following. The Josephson relations are $I=I_c \sin\phi \approx I_c \phi$ and $\partial_t \phi = \Delta\mu/\hbar$. The current dynamically changes the particle number imbalance $I = -\partial_t\Delta N/2$ which in turn changes the chemical potential imbalance $\Delta N=\kappa \Delta\mu$ via the compressibility of the reservoirs $\kappa = (\partial N/\partial\mu)_T$. Combining these expressions, we obtain the equation 
\begin{equation}
    \partial_t^2 \Delta N = -(2I_c/\hbar\kappa) \Delta N = -\omega_J^2 \Delta N
\end{equation}
which gives the Josephson frequency $\omega_J = \sqrt{2I_c/\hbar\kappa}$.

The sawtooth lattice can be projected onto the 1D region by holographically shaping an attractive, red-detuned beam with a digital micromirror device (DMD) acting as a spatial light modulator and focusing the beam through a high-resolution microscope \cite{zupancic_ultra-precise_2016}. This setup allows us to project many tightly-focused gaussian spots, each acting as a lattice site, with waists on the order of the diffraction limit (approximately $0.9\mathrm{\mu m}$), an example of which is shown in Figure~\ref{fig:sawtooth_potential}. For the simple case of a set of gaussian spots, the optimal amplitude and phase holograms can be computed analytically, while a sophisticated phase-retrieval algorithm \cite{pasienski_high-accuracy_2008} allows us to calculate the optimal holograms for more complex target potentials. Both methods are flexible enough to continuously tune the depth of the boundary sites to bring the edge states into resonance with the flat band states. For this work, a model sawtooth ladder with 3 unit cells was simulated to extract experimentally realistic parameters for the tunnelling amplitudes provided in Figure~\ref{fig:sawtooth_potential}. Eventually this will also allow us to further optimize the sawtooth lattice based on the theoretical predictions. 

The main challenge in implementing this scheme using cold atoms are the energy scales imposed by the flat band and the detection sensitivity needed to measure the critical currents. The minimum achievable temperature in the reservoirs is approximately $50 \mathrm{nK}$ which leads to a temperature broadening of the Fermi-Dirac distribution of approximately $4k_B T=4.2 \mathrm{kHz}$ and therefore limits the energy resolution of the transport spectrum to that order of magnitude. Since the transmission peaks between the dispersive and flat band states are predicted to be separated by only 600 Hz, 
transport through the flat band cannot be distinguished from transport through the dispersive bands. This means that we must either reduce the temperature or increase the tunneling rate. We can further cool the system by changing the geometry of the reservoir trap from harmonic to a uniform box-trap whose resulting de-confinement reduces the temperature \cite{su_low-temperature_2003} and would allow us to reach temperatures on the order of $30 \mathrm{nK}$. A possible method to increase the tunneling rates is to exploit a three-level system present in lithium at high magnetic fields to impose lattices with subwavelength spatial structure \cite{wang_dark_2018}. The critical currents predicted in this work are likely below the detection limit of our current experiment though it is possible that measuring the Josephson oscillation frequency instead of the critical currents directly circumvents this problem. However if the oscillations are indeed still too small to resolve, one could measure the normal dissipative transport instead of the DC Josephson effect \textit{i.e.} the response to a chemical potential imbalance rather than a phase imbalance. In this way, the strength of the signal -- the conductance through the flat band states -- can be increased simply by increasing the imposed chemical potential bias.

\section{Discussion and Conclusion}
\label{sec:discussion}

The strong effect of interactions in a flat (dispersionless) band has been predicted to lead to high critical temperatures of superconductivity~\cite{kopnin2011,heikkila2011}, as well as supercurrents and superfluidity guaranteed by quantum geometric quantities~\cite{peotta2015,liang2017}. According to theory, in a flat band the critical temperature and the superfluid weight depend linearly on the strength of the interaction that leads to Cooper pair formation; this is a direct signature of the geometric contribution of superconductivity~\cite{peotta2015,liang2017}.
The linear dependence is in striking contrast to the dispersive single band case where the critical temperature is exponentially suppressed and the superfluid weight is only weakly dependent on the interaction. Flat band superconductivity has become topical since the observation of superconductivity in twisted bilayer graphene \cite{cao2018} and other moir\'e materials \cite{shen2020,liu2020} hosting flat bands. The geometric contribution of superconductivity has been suggested to play a role there~\cite{julku2020,hu2019,xie2020,classen2020}, however, its direct verification is challenging due to the complexity of the moir\'e materials and limited possibilities of precisely tuning the interaction strength. We proposed here 
a two-terminal setup to investigate 
how the salient features of flat band superconductivity manifest in a transport experiment. Such experiment can be realized with ultracold gases where the interaction strength is highly controllable and optical lattice potentials that correspond to simple flat band models can be realized. 

We considered a finite-size sawtooth lattice in the channel between two reservoirs and characterized the DC Josephson current. We showed that the linear dependence on the interaction, as expected from the theory for infinite flat band lattice systems, is visible in the critical current. The experiment we propose would thus be able to prove the flat band nature and geometric origin of the superconductivity. The finite size of the lattice manifests itself in an intriguing way: in order to maximize the Josephson current through the bulk flat band states, one needs to make them degenerate with the boundary states that appear in a finite system and connect the lattice to the leads. Once this energy resonance condition is reached, the critical current and the critical temperature are higher in the flat band than in the dispersive bands, highlighting the general promise of flat band superconductivity. 

Our calculations used parameters obtained from microscopic modelling of real experimental potential landscapes, and we discussed the feasibility of the experiments. It is particularly important that both the interaction, and the flat band state - boundary state energy difference, can be easily controlled in the proposed ultracold gas setup. The former is needed for exploring the fundamental properties of flat band superconductivity, and the latter is useful in verifying that the effects of the finite size of the lattice are well described by the theory presented here. The main challenge in the experimental realization is the temperature scale of the current experiments, which has to be either reduced or made relatively smaller by increasing the hopping energy in the lattice.

Our results show that two-terminal transport experiments, in general, are a promising platform to explore fundamental features of flat band superconductivity, and that ultracold gas transport setups are particularly suited for this. We showed that the finite size of the flat band lattice system in the transport channel does not prevent observing the most important characteristics of flat band superconductivity, in contrast, it provides an additional turning knob for the experiments. An interesting future direction is to study also non-equilibrium flat band transport.

\acknowledgements

We acknowledge stimulating discussions with Laura Corman, Samuel Häusler and Martin Lebrat. 
V.A.J.P., S.P. and P.T. acknowledge support by the Academy of Finland under project numbers 330384, 303351, 307419, 327293, 318987 (QuantERA project RouTe), 318937
(PROFI), and by Centre for Quantum Engineering (CQE) at Aalto University. V.A.J.P. acknowledges financial support by the Jenny and Antti Wihuri Foundation. 
T.E., P.F. and J.M. acknowledge the Swiss National Science Foundation (Grants No. 182650 and No. NCCR-QSIT) and European Research Council advanced grant TransQ (Grant No. 742579) for funding.

\bibliography{main}

\end{document}